\documentclass[preprint,preprintnumbers,amsmath,amssymb,superscriptaddress]{revtex4}
\usepackage{txfonts}
\usepackage{graphicx}
\usepackage{dcolumn}
\usepackage{bm}
\usepackage{array}
\usepackage[colorlinks=true,plainpages=false,linkcolor=blue,urlcolor=blue,citecolor=blue,pdfpagemode=UseNone,pdfstartview=FitBH]{hyperref}

\begin{document}

\title{Commensurate-to-incommensurate transition of charge-density-wave order and a possible quantum critical point in pressurized kagome metal CsV$_3$Sb$_5$}

\author{X. Y. Feng}
\affiliation{Institute of Physics, Chinese Academy of Sciences,\\
	and Beijing National Laboratory for Condensed Matter Physics,Beijing 100190, China}
\affiliation{School of Physical Sciences, University of Chinese Academy of Sciences, Beijing 100190, China}

\author{Z. Zhao}
\affiliation{Institute of Physics, Chinese Academy of Sciences,\\
	and Beijing National Laboratory for Condensed Matter Physics,Beijing 100190, China}
\affiliation{School of Physical Sciences, University of Chinese Academy of Sciences, Beijing 100190, China}

\author{J. Luo}
\affiliation{Institute of Physics, Chinese Academy of Sciences,\\
	and Beijing National Laboratory for Condensed Matter Physics,Beijing 100190, China}

\author{J. Yang}
\affiliation{Institute of Physics, Chinese Academy of Sciences,\\
	and Beijing National Laboratory for Condensed Matter Physics,Beijing 100190, China}

\author{A. F. Fang}
\affiliation{Derpartment of Physics, Beijing Normal University, Beijing 100875, China}

\author{H. T. Yang}
\affiliation{Institute of Physics, Chinese Academy of Sciences,\\
	and Beijing National Laboratory for Condensed Matter Physics,Beijing 100190, China}
\affiliation{School of Physical Sciences, University of Chinese Academy of Sciences, Beijing 100190, China}
\affiliation{Songshan Lake Materials Laboratory, Dongguan, Guangdong 523808, China}

\author{H. J. Gao}
\affiliation{Institute of Physics, Chinese Academy of Sciences,\\
	and Beijing National Laboratory for Condensed Matter Physics,Beijing 100190, China}
\affiliation{School of Physical Sciences, University of Chinese Academy of Sciences, Beijing 100190, China}
\affiliation{Songshan Lake Materials Laboratory, Dongguan, Guangdong 523808, China}
\affiliation{CAS Center for Excellence in Topological Quantum Computation,\\
	University of Chinese Academy of Sciences, Beijing 100190, China}

\author{R. Zhou}
\email{rzhou@iphy.ac.cn}
\affiliation{Institute of Physics, Chinese Academy of Sciences,\\
	and Beijing National Laboratory for Condensed Matter Physics,Beijing 100190, China}
\affiliation{Songshan Lake Materials Laboratory, Dongguan, Guangdong 523808, China}

\author{Guo-qing Zheng}
\affiliation{Department of Physics, Okayama University, Okayama 700-8530, Japan}

\date{\today}

\begin{abstract}
{  Clarifying the interplay between charge density waves (CDWs) and superconductivity is important in the kagome metal CsV$_3$Sb$_5$, and pressure ($P$) can play a crucial role. Here, we present $^{121/123}$Sb nuclear quadrupole resonance (NQR) measurements under hydrostatic pressures up to 2.43 GPa in CsV$_3$Sb$_5$ single crystals. We demonstrate that the CDW gradually changes from a commensurate modulation with a star-of-David (SoD) pattern to an incommensurate one with a superimposed SoD and Tri-hexagonal (TrH) pattern stacking along the $c$-axis. Moreover, the linewidth $\delta\nu$ of $^{121/123}$Sb-NQR spectra increases with cooling down to $T_{\rm CDW}$, indicating the appearance of a short-range CDW order due to CDW fluctuations pinned by quenched disorders. The $\delta\nu$ shows a Curie-Weiss temperature dependence and tends to diverge at $P_{\rm c} \sim$  1.9 GPa, suggesting that a CDW quantum critical point (QCP) exists at $P_{\rm c}$  where $T_{\rm c}$ shows the maximum. For $P > P_{\rm c}$, spin fluctuations are enhanced when the CDW is  suppressed. Our results suggest that the maximal $T_{\rm c}$ at $P_{\rm c} \sim$ 1.9 GPa is related to the CDW QCP and the presence of spin fluctuations prevent the $T_{\rm c}$ from a rapid decrease otherwise after the CDW is completely suppressed.
}
\end{abstract}

\maketitle

\section{Introduction}
Unconventional superconductivity always arises in the vicinity of another ordered electronic state, such as a magnetic order\cite{Keimer2010}, a nematic order\cite{Chu2012}, or even a charge density wave (CDW)\cite{Wu2011}. In cuprate high-temperature superconductors, iron-pnictides or heavy-fermion compounds,  carrier dopings or externally applied pressures  can suppress the magnetic or nematic order\cite{Stewart1984,Lee2006,Stewart2011}. Quantum critical points(QCPs) and associated fluctuations were often found around the ending point of these orders and considered by many a key to understanding the mechanism of unconventional superconductivity\cite{Keimer2010,Hashimoto2012,Zhou2013,Wang2018,Luo2019}. However, unconventional superconductivity around a CDW QCP is rarely observed\cite{Gruner2017}, and whether CDW fluctuations can also mediate the electron pairing is still a mystery.

Recently, a newly discovered quasi-two-dimensional superconductor $A$V$_3$Sb$_5$ ($A$ = K, Cs, Rb) with kagome lattice has emerged as an excellent platform to study the interplay between topology, superconductivity, and CDW\cite{discovery of CsV3Sb5}. Angle-resolved photoemission spectroscopy combined with density-functional theory reveals a series of non-trivial electronic structures in this compound, including flat band, Dirac point, Van Hove singularity, and topological surface states\cite{Z2 topological kagome metal, Topological surface states and flat bands, Twofold van Hove singularity, Rich nature of Van Hove singularities, Crucial role of out-of-plane Sb p orbitals in Van Hove singularity, Dirac nodal lines and nodal loops}. Meanwhile, many exotic features, including chirality\cite{Chiral flux phase,Electronic nature of chiral charge order,Switchable chiral transport,Analysis of Charge Order}, nematicity\cite{electronic nematicity}, and time-reversal symmetry-breaking\cite{Time-reversal symmetry broken by charge order, Time-reversal symmetry breaking by polar Kerr effect}, were found in the CDW state, which was proposed to be driven by electron correlations\cite{Li2021,Possible star-of-David pattern}. Especially, the unusual phase diagram of the CDW order and superconductivity with applying hydrostatic pressures attracted a lot of attention\cite{Two types of charge order,multidome superconductivity,Double superconducting dome and triple enhancement of Tc,Unusual competition of superconductivity and charge-density-wave state}. The CDW transition can be gradually suppressed by applying hydrostatic pressure until $P_{\rm c} \sim$ 1.9 GPa, and the superconducting transition temperature $T_{\rm c}$ shows a non-monotonic double-dome-like phase diagram until its disappearance around 10 GPa\cite{Double superconducting dome and triple enhancement of Tc,Unusual competition of superconductivity and charge-density-wave state,Pressure-induced reemergence, Highly robust reentrant superconductivity}. Most remarkably, the maximum $T_{\rm c}$ is right at  $P_{\rm c}$, where no Hebel-Slichter coherence peak is seen below $T_{\rm c}$ in the superconducting state\cite{stripe-like CDW}. All these studies point to a possible QCP at $P_{\rm c}$\cite{Mechanism of exotic density wave and beyond Migdal unconventional superconductivity, a pressure-induced quantum critical point}, which makes CsV$_3$Sb$_5$ an ideal compound to study the relationship between unconventional superconductivity and CDW.
Although the high-pressure nuclear magnetic resonance(NMR) experiments suggested that the CDW undergoes an evolution to a new phase with a possible stripe-like CDW order with a unidirectional 4a$_0$ modulation in pressurized CsV$_3$Sb$_5$\cite{stripe-like CDW}, information about pressure-dependent CDW fluctuations is still lacking, which is of much significance to clarify its interaction with superconducting symmetry. Besides the CDW fluctuations, spin fluctuations were also proposed to play an important role for the high-pressure superconducting phase\cite{spin fluctuations,Mechanism of exotic density wave and beyond Migdal unconventional superconductivity}. But whether spin fluctuations exist or not and how they are affected by pressures are still unclear in the current stage.


\section{Results and Discussions}

\subsection{Commensurate-to-incommensurate transition of the CDW order}

\begin{figure}[htbp]
	\includegraphics[width=12cm]{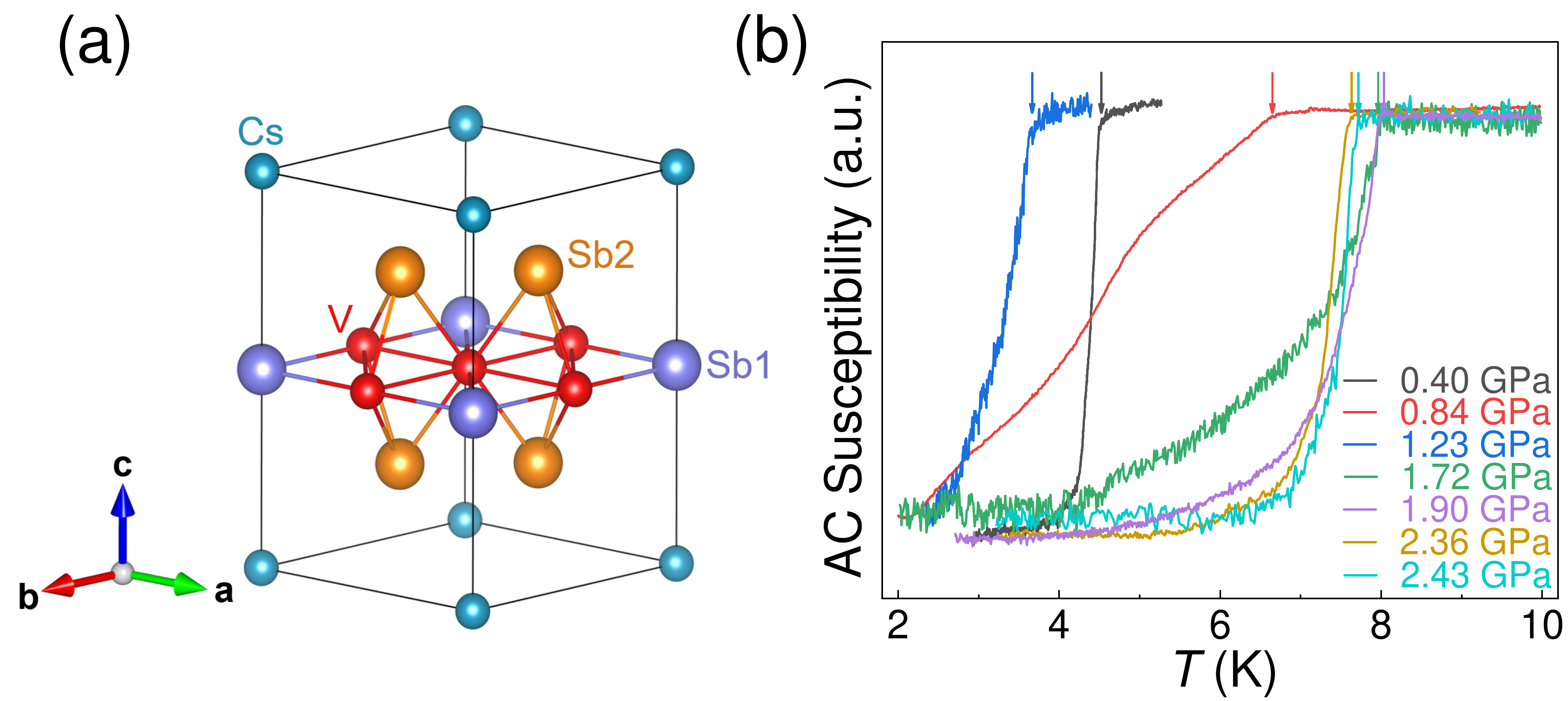}
	\centering
	\caption{(Color online) \textbf{The crystal structure and AC susceptibility measurements.} (a) The pristine crystal structure of CsV$_3$Sb$_5$ at ambient pressure. (b) The temperature dependence of the AC susceptibility measured by using an in-situ NQR coil at various pressures from 0.40 GPa to 2.43 GPa. Solid arrows represent the superconducting transition temperature $T_{\rm c}$. }
	\label{structureandtc}
\end{figure}

Figure \ref{structureandtc} shows the crystal structure and the temperature dependence of AC susceptibility measured at various pressures by using an in-situ NMR coil. The strong diamagnetic signal and the sharp superconducting transition are observed at $P = $ 0.40 GPa and $P \geq$ 1.90 GPa, indicating the high quality of the sample. As in previous studies, the much broader superconducting transitions are observed at 0.84 GPa $ \leq P \leq$ 1.72 Gpa\cite{Double superconducting dome and triple enhancement of Tc,Unusual competition of superconductivity and charge-density-wave state}.  The obtained pressure dependence of $T_{\rm c}$ is  consistent with previous transport studies(see supplementary Fig. S3)\cite{SM}.


\begin{figure}[htbp]
\includegraphics[width=16cm]{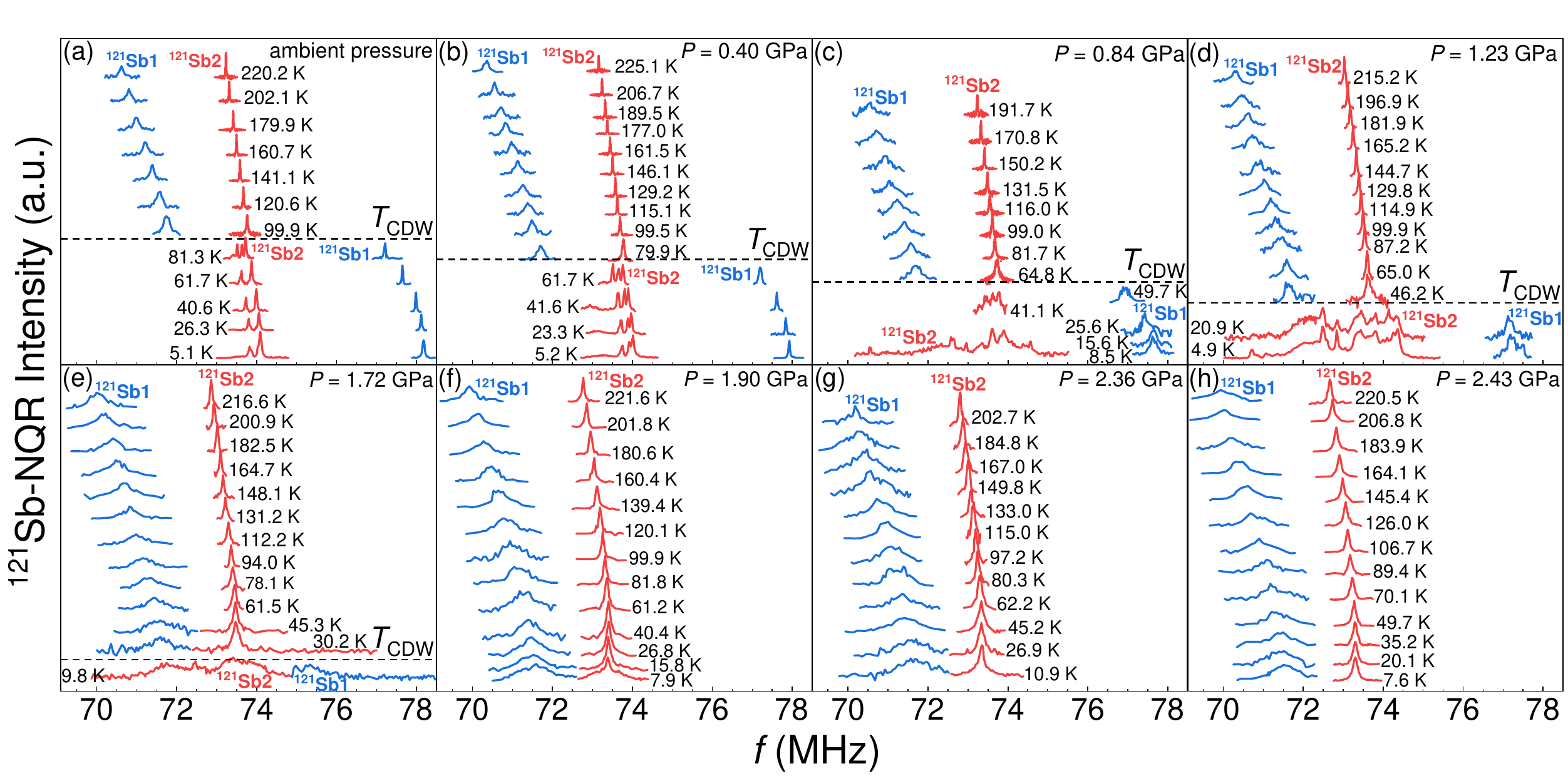}
\centering
\caption{(Color online) \textbf{Temperature-dependent NQR spectra.} The temperature dependence of $^{121}$Sb1(blue color) and $^{121}$Sb2(red color) NQR spectra at various pressures. The black dotted line indicates the temperature where the CDW phase transition occurs. }
\label{allspectra}
\end{figure}

There are two types of Sb sites in CsV$_3$Sb$_5$. Sb1 is located in the kagome plane surrounded by the vanadium hexagon, and Sb2 is located between the kagome plane and Cs layer as illustrated in Fig. \ref{structureandtc}(a).
Sb has two types of isotopes, $^{121}$Sb (\emph{I} = 5/2) and $^{123}$Sb (\emph{I} = 7/2). The quadrupole frequency $\nu_q$ is defined as $\nu_q= \frac{3e^{2}qQ}{2I(2I-1)h}$, where $eq$ is the electric field gradient (EFG) and $Q$ is the nuclear quadrupole moment. For $^{121}$Sb nucleus, the NQR spectrum should have two resonance peaks corresponding to $\pm$ 1/2 $\leftrightarrow$ $\pm$ 3/2 and $\pm$ 3/2 $\leftrightarrow$ $\pm$ 5/2 transitions. For $^{123}$Sb nucleus, the NQR spectrum should have three resonance peaks corresponding to $\pm$ 1/2 $\leftrightarrow$ $\pm$ 3/2, $\pm$ 3/2 $\leftrightarrow$ $\pm$ 5/2 and $\pm$ 5/2 $\leftrightarrow$ $\pm$ 7/2 transitions. So a total of ten lines should be observed in $^{121/123}$Sb-NQR spectrum for CsV$_3$Sb$_5$, which is indeed seen in previous NQR studies\cite{Possible star-of-David pattern, S-Wave Superconductivity}. Figure \ref{allspectra} displays the temperature dependence of $^{121}$Sb-NQR spectra corresponding to $\pm 1/2 \leftrightarrow \pm 3/2$ transitions at various pressures. For all pressures, there is only one peak for both $^{121}$Sb1 and $^{121}$Sb2 above $T_\textrm{\rm CDW}$. For $P < $1.9 GPa, a clear change of the Sb-NQR spectrum due to the CDW transition can be seen as observed at ambient pressure\cite{Possible star-of-David pattern, S-Wave Superconductivity}, but $T_{\rm CDW}$ gradually decreases with increasing pressure. The abrupt jump of Sb1 line was observed until $P$ = 1.23 GPa, indicating the CDW order is of the first order. But it is hard to determine the type of the CDW transition for $P$ = 1.72 GPa, since the line is too broad in the CDW state(see supplementary Fig. S5)\cite{SM}.

\begin{figure}[htbp]
	\includegraphics[width=9cm]{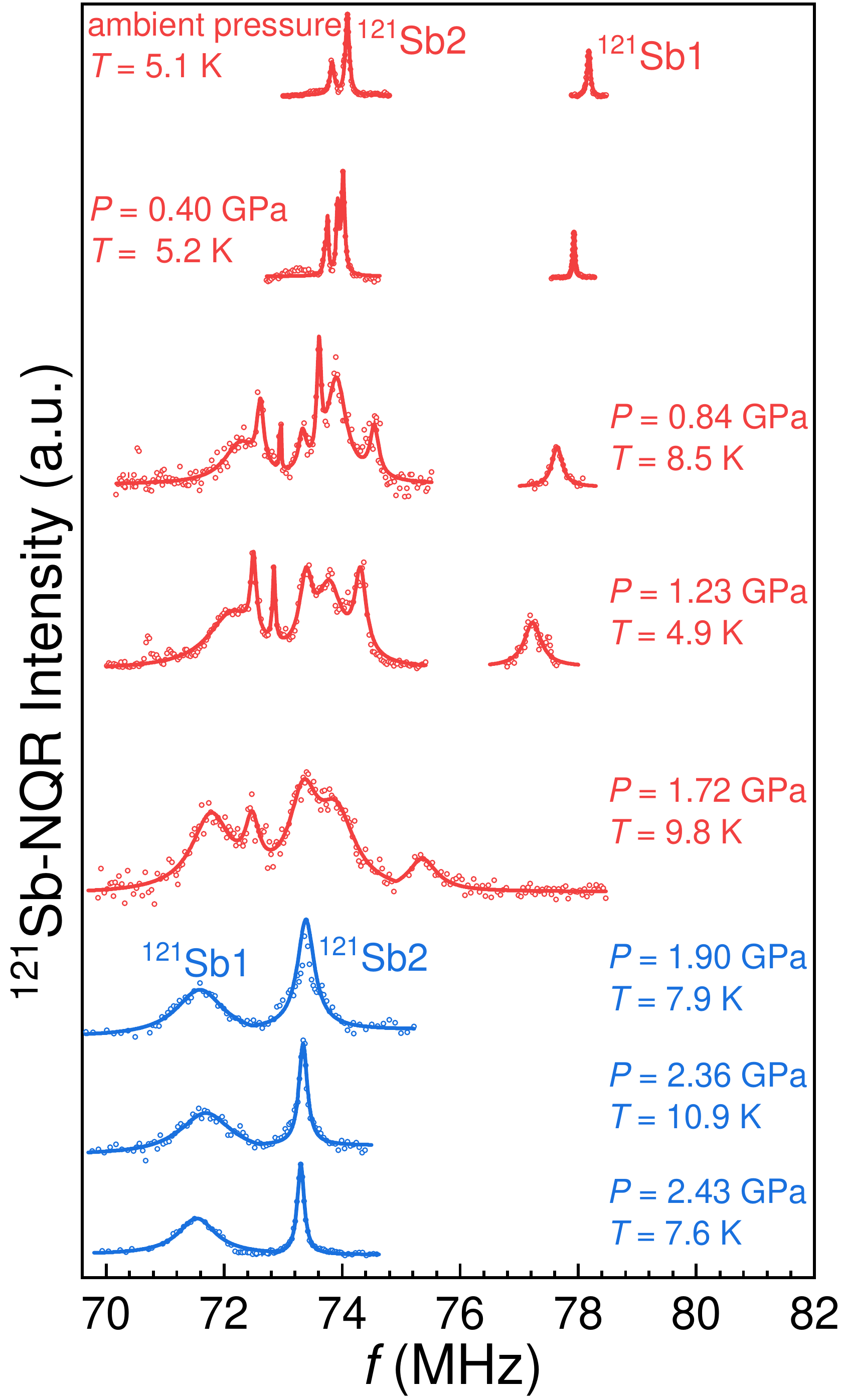}
	\centering
	\caption{(Color online) \textbf{Pressure evolution of NQR spectra.} The red peaks are $^{121}$Sb-NQR spectra in the CDW state at $T$$\sim$$T_{\rm c}$ from ambient pressure to 1.72 GPa.  The blue peaks are the $^{121}$Sb-NQR spectra at $T$$\sim$$T_{\rm c}$ above $P$ = 1.90GPa. The solid lines are the guides to the eye. }
	\label{cdw spectrum}
\end{figure}

Inside the CDW state, we further found that the lineshape of $^{121}$Sb2-NQR spectra experienced a remarkable change with increasing pressure as shown in Fig. \ref{cdw spectrum}. For $P \leq $ 0.4 Gpa, a simple splitting of $^{121}$Sb2 lines was observed, indicating that the CDW order is still commensurate and the lattice distortion should be still star-of-David (SoD) pattern\cite{Possible star-of-David pattern}. With increasing pressure, at $P$ = 0.84 GPa, both Sb1 and Sb2 NQR lines start to broaden, and new lines emerge at low frequencies. Below $P$ = 1.23 GPa, some sharp lines can still be seen between $f$ = 72 MHz to 75 MHz, but only broad lines remain at $P$ = 1.72 GPa.

\begin{figure}[htbp]
	\includegraphics[width=17cm]{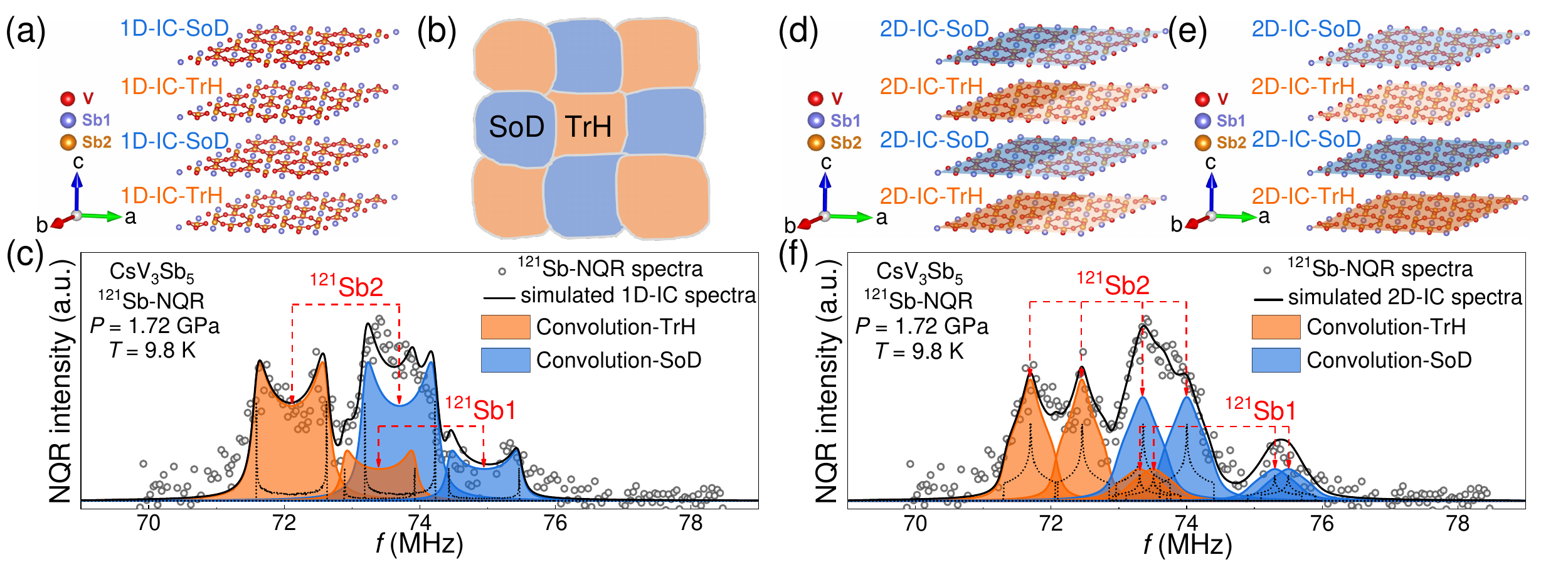}
	\centering
	\caption{(Color online) \textbf{Simulation of $^{121}$Sb-NQR spectrum at $P$ = 1.72 GPa.} (a) and (b) display the two possible CDW patterns with an one-dimensional(1D) incommensurate(IC) modulation, in which (a) represents the superimposed SoD and TrH pattern stacking along the c-axis and (b) represents the coexistence of SoD and TrH domains, respectively. (d) and (e) illustrate the superimposed two-dimensional(2D) incommensurate SoD and TrH pattern with an additional charge modulation along the $a$-axis and $c$-axis, respectively. The gray level represents the charge density. (c) and (f) show the comparison of the $^{121}$Sb-NQR spectra at $P$ = 1.72 GPa (grey circle), the simulated incommensurate spectra (black dotted line) and the calculated convolution (orange area for TrH pattern and blue area for SoD pattern) for 1D and 2D incommensurate CDW modulations(details about NQR spectra simulation are present in Supplementary Note 5)\cite{SM}, respectively. The peaks corresponding to Sb1 and Sb2 sites are marked by the red dashed arrows. }
	\label{simulation}
\end{figure}

The observed broadening and emergence of new lines imply that the CDW modulation is totally different from the modulation at ambient pressure. Below we show that an incommensurate(IC) CDW order with superimposed SoD and TrH pattern stacking along the $c$-axis can consistently account for the observed results. Generally speaking, in a commensurate CDW state, the NQR line reflects the small number of physically non-equivalent nuclear sites in the unit cell so that the spectrum with discrete peaks was observed. In an incommensurate state, however, since the translational periodicity is lost, the number of non-equivalent nuclear sites becomes much larger and leads to a larger broadening\cite{Blinc}. A modulation due to CDW order will cause an additional term in the resonance frequency at Sb site ($x,y$). In our model, we consider both one-dimensional (1D) and two-dimensional (2D) incommensurate modulations. In the 1D case,  we assume that the charge modulation along one in-plane direction is incommensurate and introduce an additional cosine function as $\cos \left( \frac{2\pi }{a}{{q}_{x}}\cdot x \right)$\cite{Blinc}. In the 2D case,  we assume that the incommensurate modulation is in-plane and introduce an additional term be $\cos \left ( \frac{2\pi}{a}  q_x\cdot x \right ) + \cos \left [ \frac{2\pi}{b} \left (  q_y\cdot x\cdot \cos \beta + q_y\cdot y\cdot \sin \beta  \right ) \right ]$\cite{Blinc}. $q_x$ and $q_y$ are the wave vectors along $a$ and $b$-axis, respectively. $\beta$ is the angle between the two in-plane wave vectors $q_x$ and $q_y$, which is $\pi/3$ for the kagome lattice studied in this work.  For the 1D incommensurate case, we propose that the SoD and TrH patterns could be either  superimposed as illustrated in Fig. \ref{simulation}(a) or formed two different domains as illustrated in Fig. \ref{simulation}(b).
For the 2D incommensurate case, we assume an additional charge modulation on top of the superimposed SoD and TrH pattern, either along the $a$-axis or $c$-axis as illustrated in Fig. \ref{simulation} (d) and (e). By only considering the structural distortion in the plane and convoluting with a Lorentz function (details about NQR spectra simulation are present in Supplementary Note 5)\cite{SM}, we can reproduce the spectra at $P$ = 1.72 GPa for both 1D and 2D incommensurate modulation  as shown in Fig. \ref{simulation}(c) and (f), respectively.

However, for the 1D incommensurate modulation, the Sb1 NQR spectra should have two peaks of equal intensities at 74.5 and 75.4 MHz, respectively (see Fig. \ref{simulation}(c)), which is not observed at $P$ = 0.84 GPa or 1.23 GPa (see Fig. \ref{cdw spectrum}).
We note that a stripe CDW order was proposed by the previous $^{51}$V-NMR study\cite{stripe-like CDW}, which is similar to our assumption of the additional modulation along the $a$-axis. However, the incommensurability of CDW and the coexistence of SoD and TrH patterns were not caught by the $^{51}$V-NMR. This might be because Sb nuclei are sensitive to charge modulation from the Sb 5$p$-orbitals, which was suggested to be different from the CDW originated from the V 3$d$-orbitals\cite{Li20222}. In addition, $^{121/123}$Sb-NQR spectra were found to have a much larger response to the CDW order compared to the $^{51}$V-NMR spectra\cite{Possible star-of-David pattern,S-Wave Superconductivity}. In any case, our results suggest that CDW modulation gradually changes from the commensurate CDW at ambient pressure to the incommensurate CDW with increasing pressure. However, the NMR line shape is independent of the value of the CDW
wave vector $q$ for incommensurate modulations. Moreover, the present experimental results do not rule out the possibility of more complex CDW patterns beyond the proposed structures in Fig. \ref{simulation}. To further resolve this issue, high-pressure X-ray scattering measurements at 1.72 GPa are needed in the future.

In the range of   0.84GPa $ \leq P \leq $ 1.23 GPa, we found that the NQR spectra consist of both narrow and broad peaks(see Supplementary Fig. S7)\cite{SM}, indicating the coexistence of the commensurate and incommensurate CDWs. Then, there will be a large number of CDW domain walls between the commensurate and incommensurate CDWs in this pressure region. The enhanced interaction and scattering at the domain walls can strongly affect the superconductivity\cite{Unusual competition of superconductivity and charge-density-wave state,Lee2021}, which is likely responsible for the inhomogeneous superconductivity as we found (see Supplementary Fig. S8 for the comparison between the commensurate CDW volume fraction and the superconducting transition width)\cite{SM}.

A commensurate to incommensurate transition with increasing pressures as we found was recently proposed theoretically \cite{Mechanism of exotic density wave and beyond Migdal unconventional superconductivity}, but a superimposed SoD and TrH pattern was not predicted. In CsV$_3$Sb$_5$, instead of electron-phonon coupling, electron correlations were suggested to be an important factor in forming the CDW order\cite{Li2021}.  Most interestingly, the incommensurate modulation was also reported in Sn-doped CsV$_3$Sb$_5$\cite{Kautzsch2022}. Then, one possible scenario is that the ordering wave vector connects parts of the Fermi surface or the hot spots. And with increasing pressure, due to the change of the Fermi surface, the ordering wave vector gradually becomes incommensurate.
Such a scenario was proposed for the CDW order in cuprates, and the wave vector was found to have a monotonous doping-dependence\cite{Comin2016}. It would be interesting to measure the doping dependence of the wave vector by the high-pressure X-ray scattering.

\subsection{Possible CDW quantum critical point}

\begin{figure}[htbp]
	\includegraphics[width=17cm]{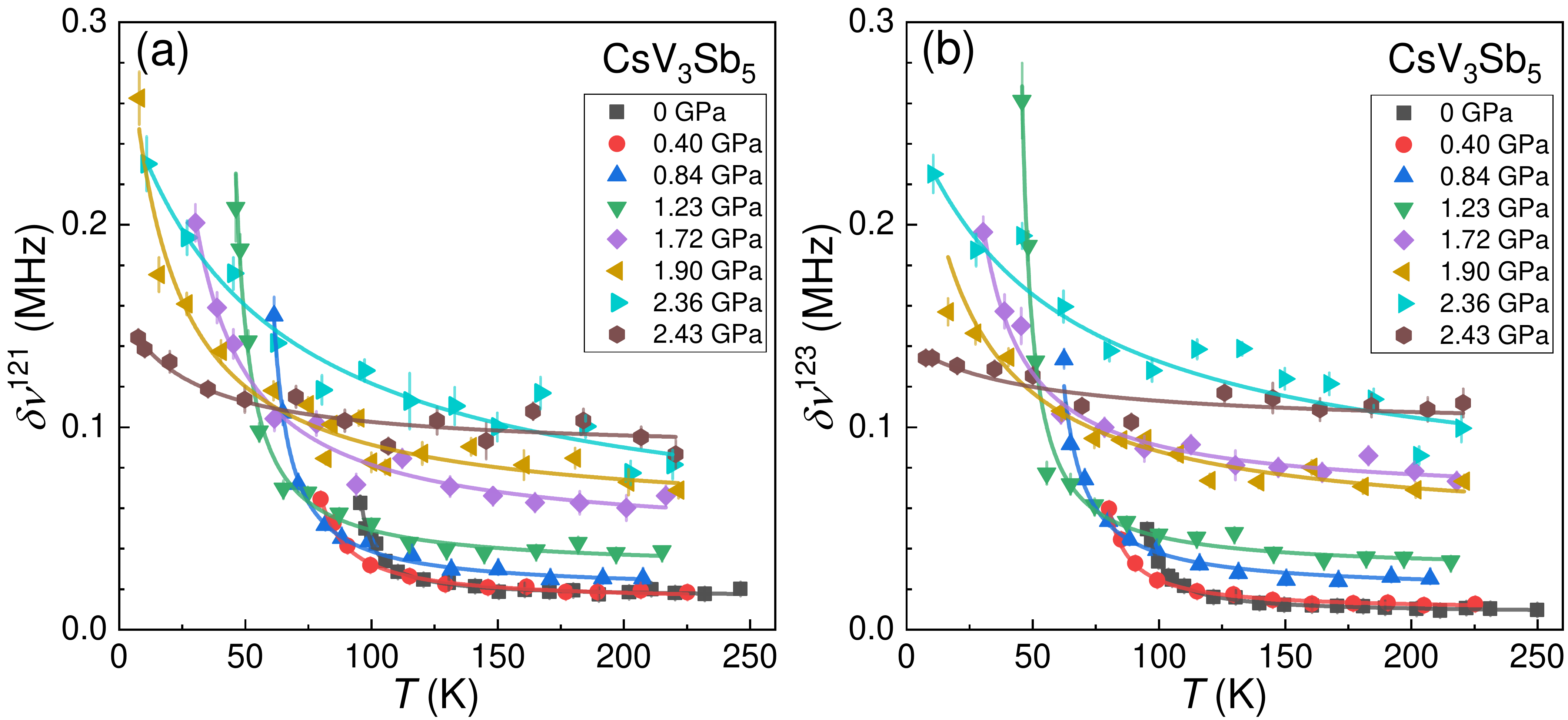}
	\centering
	\caption{(Color online) \textbf{The NQR linewidth.} The full-width at half-maximum (FWHM) $\delta\nu$ of $^{121/123}$Sb2 NQR spectra at various pressures. The solid lines are Curie-Weiss fits and the obtained $\theta$ values are plotted in Fig. \ref{phasediagram}. Error bars are s.d. in the fits of the NQR spectra. }
	\label{fwhm}
\end{figure}

\begin{figure}[ht]
	\includegraphics[width=12cm]{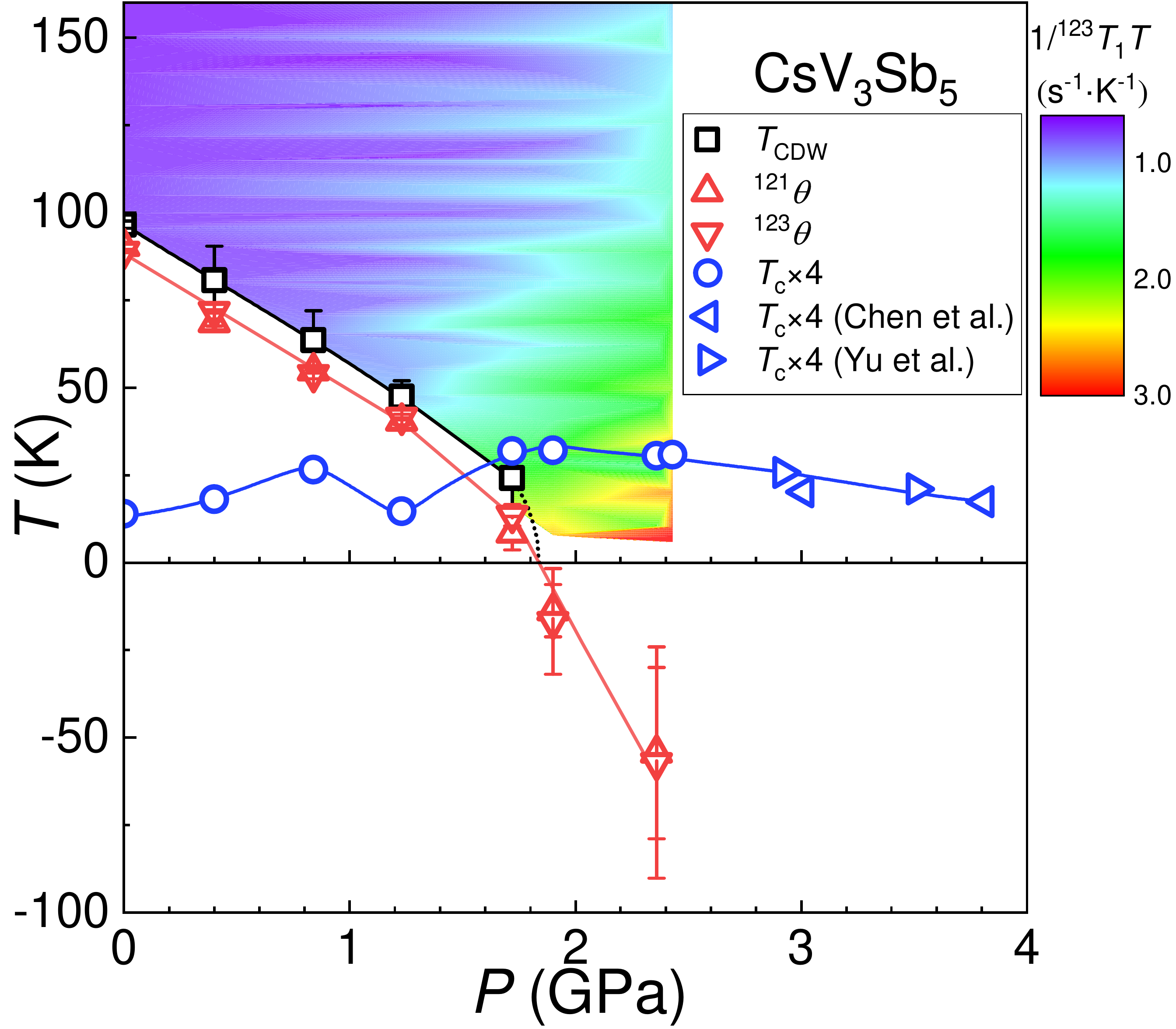}
	\centering
	\caption{(Color online) \textbf{ The obtained phase diagram of CsV$_3$Sb$_5$.} The black square is the CDW transition temperature $T_{\rm CDW}$ determined by the temperature-dependence of Sb2 NQR spectral intensity (see Supplementary Fig. S4)\cite{SM}. The blue circle is superconducting transition temperature $T_{\rm c} \times 4$ obtained in this work. The blue triangle is $T_{\rm c} \times 4$ taken from previous transport measurements\cite{Double superconducting dome and triple enhancement of Tc,Unusual competition of superconductivity and charge-density-wave state}. The red triangle is the obtained $^{121/123}$$\theta$ from the Curie-Weiss fitting in Fig. \ref{fwhm}. For $P$ = 2.43 GPa, $\delta\nu$ has a very weak temperature dependence, which leads to a large error bar $\sim$ 100 K from the Curie-Weiss fitting. So we did not plot $^{121/123}$$\theta$ at $P$ = 2.43 GPa in the phase diagram. Colours in the normal state represent the evolution of the  $1/T_1T$ of $^{123}$Sb2. Solid and dashed lines are guides to the eye. The error bar for $T_{\rm CDW}$ represents the temperature interval in measuring the NQR spectra (see Supplementary Fig. S4)\cite{SM}. }
	\label{phasediagram}
\end{figure}

Next, we turn to the fluctuations above $T_{\rm CDW}$. By fitting the $^{121}$Sb2 and $^{123}$Sb2 spectra with the Lorentz function (see Supplementary Fig. S9 for $^{123}$Sb-NQR spectra of the Sb2 site)\cite{SM}, we deduced the linewidth $\delta$$\nu^{121}$ and $\delta$$\nu^{123}$ at various pressures as shown in Fig. \ref{fwhm}(a) and (b), respectively. Both $\delta$$\nu^{121}$ and $\delta$$\nu^{123}$ increases with decreasing temperature until $T \sim T_{\rm CDW}$, indicating the existence of the short-range CDW order due to CDW fluctuations pinned by quenched disorders, which was also observed in 2H-NbSe$_2$ and underdoped cuprate YBa$_2$Cu$_3$O$_y$\cite{Berthier1978,Ghoshray2009,Wu2015}.
Our observation is consistent with the recent X-ray scattering and specific heat measurements at ambient pressure, which also show the existence of a short-range CDW order above $T_{\rm CDW}$\cite{Chen2022}.
Moreover, we find that the temperature dependence of $\delta$$\nu$ also follows the Curie-Weiss behavior as observed in YBa$_2$Cu$_3$O$_y$\cite{Vinograd2019}, and fit both $\delta$$\nu^{121}$ and $\delta$$\nu^{123}$ by the Curie-Weiss formula as,
\begin{equation}
	\delta\nu^{121,123}(T) = \frac{A^{121,123}}{T-\theta^{121,123}} + C^{121,123}
\end{equation}	
where $A$ represents the amplitude of the Curie-Weiss fit and $C$ is a constant. As shown in Fig. \ref{fwhm}, both $\delta$$\nu^{121}$ and $\delta$$\nu^{123}$ are fitted very well, and the obtained $^{121}\theta$ and $^{123}\theta$  are plotted in the phase diagram(see red triangles in Fig. \ref{phasediagram}). Most surprisingly, we find that both $^{121}\theta$ and $^{123}\theta$ are very close to $T_{\rm CDW}$ from the ambient pressure to $P$ = 1.72 GPa, indicating a divergent behavior of $\delta$$\nu$. Therefore, our results suggest that the NQR line broadening approaching $T_{\rm CDW}$ is related to the CDW fluctuations.
There is no present theory giving the quantitative relationship between the CDW susceptibility and the NQR line width $\delta$$\nu$, however, in analogy with the magnetic and nematic quantum phase transitions\cite{Hashimoto2012,Zhou2013},  we can take $\theta$ as an indicator of the QCP. If $\theta$  = 0, it means that the CDW susceptibility diverges at $T \rightarrow 0$, indicating a CDW QCP. As shown in Fig. \ref{phasediagram}, both $^{121}\theta$ and $^{123}\theta$ are almost zero at $P_{\rm c} \sim$ 1.9 GPa, suggesting a CDW QCP at this pressure. We note that $T_{\rm c}$ reaches the maximum at $P_{\rm c}$, implying the possible relationship between CDW fluctuations and the superconductivity. In order to make a firm conclusion, it will be important to make sure whether the CDW transition is of second-order at $P >$  1.72GPa, and whether the CDW QCP is beneath the superconducting dome\cite{Shibauchi2014,Wang2018}.



\subsection{Spin fluctuations}

\begin{figure}[htbp]
	\includegraphics[width=16cm]{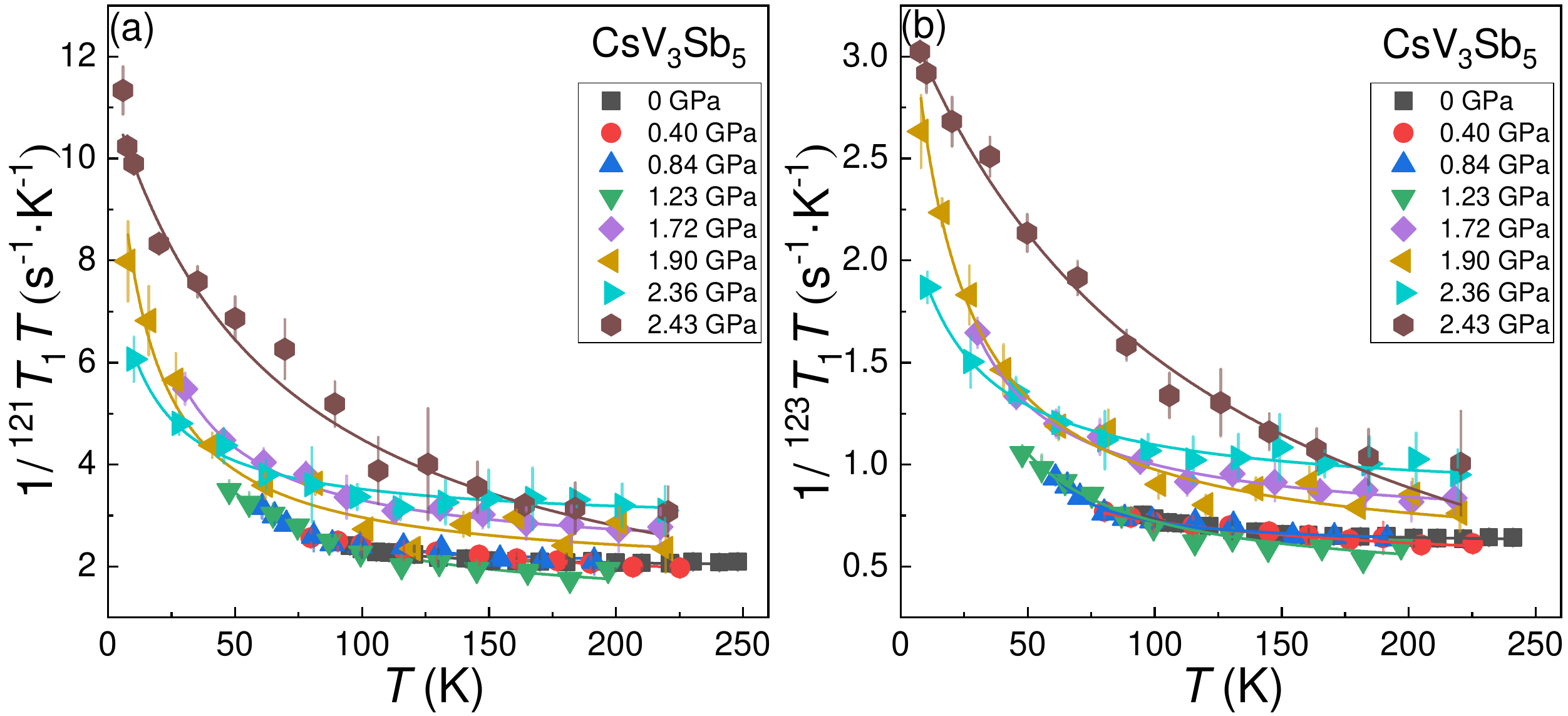}
	\centering
	\caption{(Color online) \textbf{The quantity 1/$T_1T$.} (a) and (b) are temperature-dependent $1/^{121}T_1T$ and $1/^{123}T_1T$ measured at Sb2 site under various pressures. Solid lines are guides to the eye. Error bars are s.d. in the fits of the nuclear magnetization recovery curve. }
	\label{1_T1T}
\end{figure}

\begin{figure}[ht]
	\includegraphics[width=10cm]{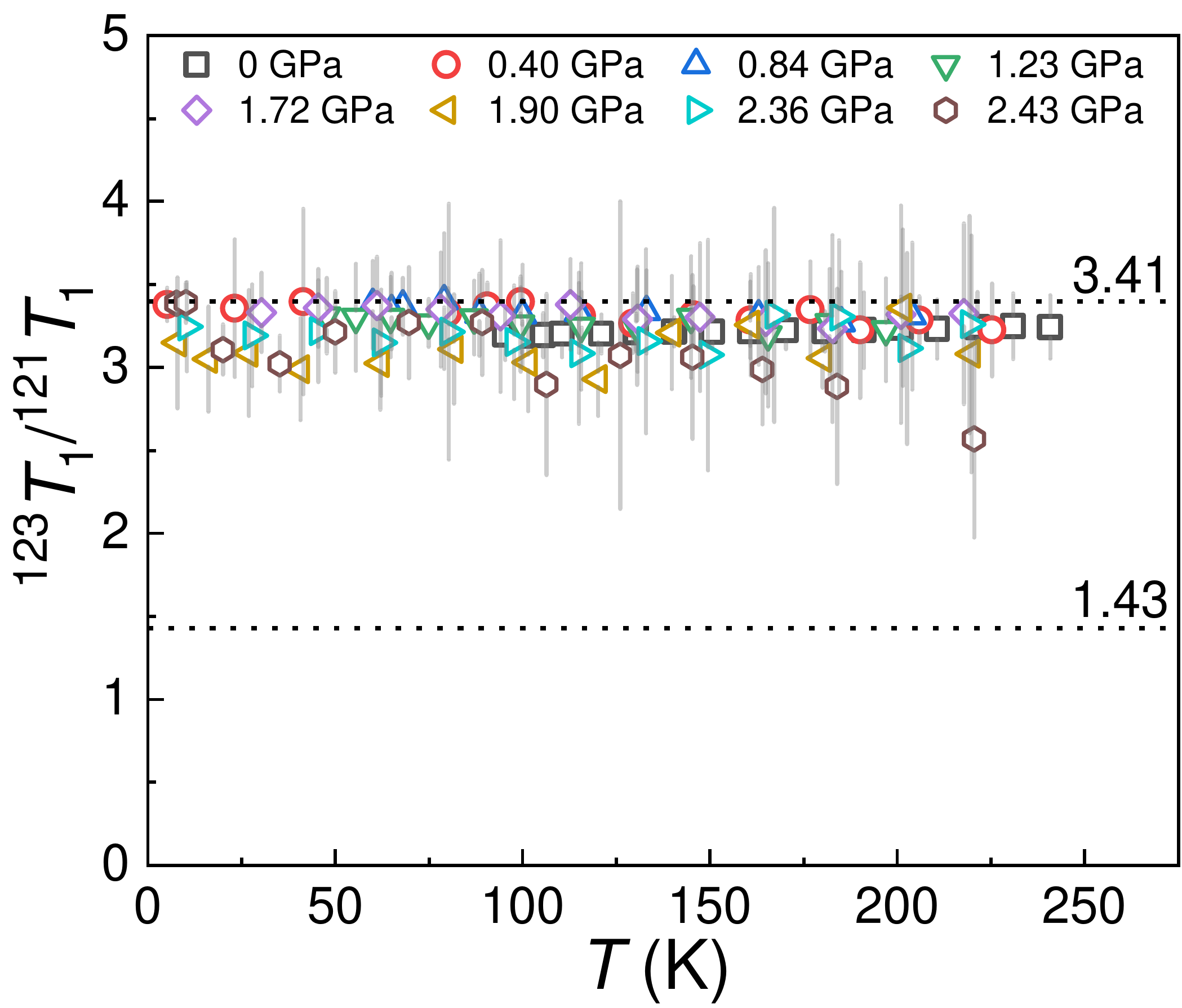}
	\centering
	\caption{(Color online) \textbf{Temperature dependence of the 1/$T_1$ ratio $^{123}$$T_1$/$^{121}$$T_1$.} The horizontal dashed lines represent purely magnetic fluctuations (ratio = 3.41) and EFG fluctuations (ratio = 1.43), respectively. Error bars are s.d. in the fits of the nuclear magnetization recovery curve.}
	\label{fluctuationsource}
\end{figure}

Lastly, we tried to obtain more information about fluctuations by measuring the spin-lattice relaxation rate 1/$T_1$ at both $^{121}$Sb2 and $^{123}$Sb2 sites at various pressures as shown in Fig. \ref{1_T1T}. At all pressures, $1/T_1T$ increases with decreasing temperature towards $T_{\rm CDW}$. To further show the evolution of 1/$T_1T$, we make a contour plot in Fig. \ref{phasediagram}. 1/$T_1T$ is almost identical for $P <$ 1.72 GPa, but starts to be enhanced from $P$ = 1.9 GPa after the fully suppression of the CDW order, which shows a totally different behavior comparing to the NQR line broadening (see Fig. \ref{fwhm}). The nuclear spin-lattice relaxation rate 1/$T_1$ is mainly composed of two contributions including magnetic interaction and quadrupole interaction. If the quadrupole relaxation process is predominant, the 1/$T_1$ ratio between $^{121}$Sb and $^{123}$Sb is expected to be $[^{121}Q^2(2\cdot^{121}I +3)/^{121}I^2(2\cdot^{121}I - 1)]/[^{123}Q^2(2\cdot^{123}I +3)/^{123}I^2(2\cdot^{123}I - 1)] = 1.43$\cite{Ishida}, in which $^{121}Q$ = $-0.53\times10^{-24}$ cm$^2$ and $^{123}Q$ = $-0.68\times10^{-24}$ cm$^2$ are taken. If the magnetic relaxation process is predominant, the 1/$T_1$ ratio between $^{121}$Sb and $^{123}$Sb is expected to be $(^{121}\gamma/^{123}\gamma)^2 = 3.41$, in whcih $^{121}\gamma$ = 10.189 MHz/T and $^{123}\gamma$ = 5.51756 MHz/T are taken. As shown in Fig. \ref{fluctuationsource}, the 1/$T_1$ ratio $^{123}T_1$/$^{121}T_1$ is close to 3.41 for all pressures, indicating that $1/T_1T$ is mainly contributed by spin fluctuations. Therefore, our results suggest the existence of spin correlations in CsV$_3$Sb$_5$. With increasing pressure, the spin correlations are significantly enhanced after the complete suppression of CDW order.
More interestingly, as reported by previous transport studies, $T_{\rm c}$ does not drop rapidly for $P >$ 1.9 GPa(see Supplementary Fig. S10 for the complete phase diagram)\cite{Double superconducting dome and triple enhancement of Tc,Unusual competition of superconductivity and charge-density-wave state,SM}. Our results suggest that the superconductivity is sustained by the spin fluctuations at high pressures, which seems to be consistent with recent theoretical studies\cite{spin fluctuations,Mechanism of exotic density wave and beyond Migdal unconventional superconductivity}. In passing, we also note that a new superconducting state arises above $P \sim$ 15 GPa with the pressure further increasing\cite{Pressure-induced reemergence, Highly robust reentrant superconductivity}. Whether spin fluctuations still play a role for such a higher pressure phase needs high-pressure NMR measurements by using diamond anvils to clarify.

\section{Conclusions}

In conclusion, we have presented the systematic $^{121/123}$Sb NQR measurements on CsV$_3$Sb$_5$ under hydrostatic pressures. We found that the CDW structure gradually changes from a commensurate SoD pattern at ambient pressure to a superimposed incommensurate SoD and TrH pattern at $P$ = 1.72 GPa. Above $T_{\rm CDW}$, we find that the linewidth of NQR spectra increases with decreasing temperature, indicating the existence of CDW fluctuations pinned by quenched disorders. The linewidth shows a Curie-Weiss temperature dependence and tends to diverge at $P_c \sim$  1.9 GPa where $T_{\rm c}$ shows the maximum. Spin fluctuations are enhanced for $P \geqslant P_c$, which is probably responsible for the slow decrease of $T_{\rm c}$ at high pressures.  Our results reveal the evolution of CDW structure and an emerged CDW QCP with increasing hydrostatic pressures, providing new insight into the superconducting pairing mechanism in CsV$_3$Sb$_5$.

\vspace{0.5cm}

\textbf{Methods}

\textbf{Sample preparation and NQR measurement}

Single crystal CsV$_3$Sb$_5$ was synthesized by the self-flux method\cite{discovery of CsV3Sb5}. The typical size of the single crystal is around 3 mm $\times$ 2 mm $\times$ 0.1 mm. Several single crystals were mounted inside a piston-cylinder pressure-cell made of CuBe alloy. To maintain consistency and ensure the number of quenched disorders remains unchanged, all measurements were conducted on the same single crystals.

The NQR measurements were performed with a phase-coherent pulsed NQR spectrometer. The $^{121/123}$Sb-NQR spectra were acquired by sweeping the frequency point by point and integrating spin-echo signal. Since the EFG principal axis of $^{121/123}$Sb is along the $c$-axis\cite{Possible star-of-David pattern}, we stack the CsV$_3$Sb$_5$ single-crystal flakes along the $c$ direction to obtain better NQR signal. The nuclear spin-lattice relaxation rate 1/$T_1$ was measured by the saturation-recovery method. The $^{121}T_1$ was obtained by fitting the nuclear magnetization $M(t)$ with $1-M(t)/M(0)=\frac{3}{28}\textrm{exp}(-3t/T_1)+\frac{25}{28}\textrm{exp}(-10t/T_1)$ and $^{123}T_1$ was fitted by $1-M(t)/M(0)=\frac{9}{97}\textrm{exp}(-3t/T_1)+\frac{16}{97}\textrm{exp}(-10t/T_1)+\frac{72}{97}\textrm{exp}(-21t/T_1)$, where $M(0)$and $M(t)$ are the nuclear magnetization respectively at thermal equilibrium and time $t$ after the comb pulse.

\textbf{High Pressure NQR measurement}

We used a commercial BeCu/NiCrAl clamp cell from C$\&$T Factory Co., Ltd. (Japan) and Daphne oil 7373 as a transmitting medium\cite{Yokogawa2007}. When we applied the pressure above 1.7 GPa, we heated the pressure cell up to 315 K to prevent the solidification of the pressure medium Daphne 7373\cite{Yokogawa2007}. Although cares have been taken, there is still a possibility that the pressure might be uniaxial at higher pressures, which could broaden the NQR lines at high temperatures as shown in Fig. \ref{fwhm}.  For 0.4 GPa $\leq P \leq$ 2.36 GPa, the applied pressure has been calibrated by the NQR frequency $^{63}\nu_Q$ of Cu$_2$O\cite{Cuprous oxide manometer, Space efficient opposed-anvil high-pressure cell}. The Cu$_2$O powder and single crystal CsV$_3$Sb$_5$ were placed together inside the NQR coil. There is a pressure deficit from room temperature to low temperature due to the solidification of Daphne oil 7373\cite{Yokogawa2007}, so the pressure-cell was pressurized at room temperature and the NQR frequency of $^{63}$v$_Q$ was measured at  $T$ $\sim$ 5 K(see Supplementary Fig. S1)\cite{SM}. The $\nu_q$ of Sb2 shows a linear pressure dependence(see Supplementary Fig. S2 (d))\cite{SM}. For $P$ = 2.43 GPa, the applied pressure was obtained by the value of $\nu_q$ at $T$ = 100 K.


\vspace{0.5cm}
\textbf{Data Availability}

The data that support the findings of this study are available from the corresponding authors upon reasonable request.

\begin{acknowledgments}
This work was supported by the National Natural Science Foundation of China (Grant Nos. 11974405, Nos. 11674377, Nos. 11634015, Nos. 51771224, Nos. 61888102 and Nos. 11904023), the National Key Research and Development Projects of China (Grant Nos. 2022YFA1403400, Nos. 2018YFA0305800 and 2019YFA0308500) and the Strategic Priority Research Program of the Chinese Academy of Sciences (Grant Nos. XDB33010100 and Nos. XDB33030100).
\end{acknowledgments}


\vspace{0.5cm}
\textbf{Author contributions}

The single crystals were grown by Z.Z., H.T.Y. and H.J.G. The NMR measurements were performed by X.Y.F., J.L., J.Y., A.F.F. and R.Z. R.Z. and G.Q.Z. wrote the manuscript with inputs from X.Y.Feng. All authors have discussed the results and the interpretation.

\vspace{0.5cm}

\textbf{Competing Interests}

The authors declare no competing interests.

\end{document}